# Enhanced Linear-array Photoacoustic Beamforming using Modified Coherence Factor

**Moein Mozaffarzadeh**[a,c], **Yan Yan**[d], **Mohammad Mehrmohammadi**[d], **Bahador Makkiabadi**[a,b,*]

[a]Research Center for Biomedical Technologies and Robotics (RCBTR), Institute for Advanced Medical Technologies (IAMT), Tehran, Iran.
[b]Department of Medical Physics and Biomedical Engineering, School of Medicine, Tehran University of Medical Sciences, Tehran, Iran.
[c]Department of Biomedical Engineering, Tarbiat Modares University, Tehran, Iran.
[d]Department of Biomedical Engineering, Wayne State University, Detroit, MI, USA.

**Abstract.** Photoacoustic imaging (PAI) is a promising medical imaging modality providing the spatial resolution of ultrasound (US) imaging and the contrast of optical imaging. For linear-array PAI, an image beamformer can be used as the reconstruction algorithm. Delay-and-sum (DAS) is the most prevalent beamforming algorithm in PAI. However, using DAS beamformer leads to low resolution images as well as high sidelobes due to non desired contribution of off-axis signals. Coherence factor (CF) is a weighting method in which each pixel of the reconstructed image is weighted, based on the spatial spectrum of the aperture, to mainly improve the contrast. In this paper, we demonstrate that the numerator of the formula of CF contains a DAS algebra, and it is proposed to use the delay-multiply-and-sum (DMAS) beamformer instead of the available DAS on the numerator. The proposed weighting technique, modified CF (MCF), has been evaluated numerically and experimentally compared to CF. It was shown that MCF leads to lower sidelobes and better detectable targets. The quantitative results of the experiment (using wire targets) show that MCF leads to for about 45% and 40% improvement, in comparison with CF, in the terms of signal-to-noise ratio and full-width-half-maximum, respectively.

*Bahador Makkiabadi, b-makkiabadi@tums.ac.ir

## 1 Introduction

Photoacoustic imaging (PAI), also called optoacoustic imaging, is an emerging medical imaging technique which combines the properties of optical and ultrasound (US) imaging.[1,2] PAI is based on the photoacoustic (PA) effect, and combined US and PA properties provide structural, functional and potentially the molecular information of tissue.[3,4] In this imaging modality, acoustic waves are generated, as a result of an electromagnetic pulse illumination, based on thermoelastic effect.[5] Then, the optical absorption distribution map of the tissue is reconstructed through a reconstruction algorithm.[6] PAI is a scalable imaging modality used in different preclinical and clinical applications e.g., tumor detection,[7,8] ocular imaging,[9] monitoring oxygenation in blood vessels,[10] and functional imaging.[5,11] There are two types of PAI: photoacoustic tomography (PAT) and photoacoustic microscopy (PAM).[12–14] In PAT, an array of US transducers in the form of linear, arc or circular shape is used for data acquisition, and mathematical reconstruction algorithms are used to obtain optical absorption distribution map of the tissue.[15] Recently, low-cost PAT and PAM systems are extensively being investigated[16–19]

In linear-array PAI, image reconstruction is done with beamformers, as in US imaging. The problem of image reconstruction in linear-array imaging for PAI and US imaging can be addressed in almost a same way. There are some modifications which should be considered in image reconstruction for these two imaging modalities, and the modifications are directly concerned with the transmission part.[20] In US imaging, US pulses are transmitted, but in PAI, the laser illumination plays the excitation role. There are many studies focused on using one beamforming technique for US and PA image formation to reduce the cost of the integrated US/PA system.[21–23] Delay-and-sum (DAS), as the most basic and commonly used beamformer in US and PAI due to its simple implementation, is a blind beamformer and results in low quality images.[24] Development of a proper beamforming algorithm has been widely investigated in US imaging in different studies.[25–28] Adaptive beamforming such as minimum variance (MV) can be a proper option to weight the signals and reduce the effect of the off-axis signals in the reconstructed images.[29] MV combined with CF has been used for PAI.[30] Short-lag spatial coherence (SLSC) beamformer was used in PAI for contract enhancement.[31] Recently, to address the relatively poor appearance of interventional devices such as needles, guide wires, and catheters, in conventional US images, delay and standard deviation (DASD) beamforming algorithm was introduced.[32] In 2015,[33] Matrone *et al*. introduced a new beamforming algorithm namely delay-multiply-and-sum (DMAS). This algorithm was initially used as a reconstruction algorithm in confocal microwave imaging for breast cancer detection.[34] Although it leads to a higher resolution compared to DAS, the resolution is not well enough in comparison with the resolution gained by MV-based algorithms. MV beamformer has been combined with DMAS algorithm to improve the resolution of DMAS.[35,36] Double stage DMAS (DS-DMAS) was introduced for PAI.[37,38] In addition, it was shown that it outperforms DMAS in the terms of contrast and sidelobes for US imaging too.[39] Eigenspace-Based Minimum Variance (EIBMV) and forward-backward (FB) MV beamformers also have been applied to medical US imaging to improve the image quality and robustness.[40,41] EIBMV was combined with DMAS to further improve the PA image quality.[42,43]

In this paper, a novel version of coherence factor (CF) algorithms is introduced. We have demonstrated that the numerator in the formula of the CF weighting procedure is the output of DAS algorithm, and it is proposed to improve the image quality by including DMAS algebra in CF, instead of the existing DAS.

The rest of the paper is organized as follows. Section 2 contains the theory of beamformers and the proposed method. Numerical simulation of the imaging system and the experimental design along with the results, and the performance evaluation are presented in section 3 and section 4, respectively. Discussion is presented in section 5, and finally the conclusion is presented in section 6.

## 2 Materials and Methods

PA signals are generated and detected after the laser has illuminated the imaging target. The obtained signals can be used to reconstruct the PA images through a reconstruction algorithm such as DAS which can be written as follows:

$$y_{DAS}(k) = \sum_{i=1}^{M} x_i(k - \Delta_i), \tag{1}$$

where $y_{DAS}(k)$ is the output of beamformer, $k$ is the time index, $M$ is the number of array elements and $x_i(k)$ and $\Delta_i$ are the detected signals and the corresponding time delay for detector $i$, respectively.[44] To provide a more efficient beamformer and improve the quality of the reconstructed image, coherence factor (CF) can be used combined with DAS, which leads to sidelobe levels reduction and contrast enhancement.[45] CF, as a weighting procedure, is presented by:

$$CF(k) = \frac{\left|\sum_{i=1}^{M} x_{id}(k)\right|^2}{M \sum_{i=1}^{M} |x_{id}(k)|^2}, \tag{2}$$

where $x_{id}$ is the delayed detected signal. The output of combined DAS and CF is given by:

$$y_{DAS+CF}(k) = CF(k) \times y_{DAS}(k). \tag{3}$$

Implementing the DAS beamformer is simple which is why is it the most common beamforming algorithm in US and PAI. However, this algorithm provides a low off-axis signal rejection and noise suppression. Consequently, DAS results in reconstructed images having high levels of sidelobe and a low resolution. To address the limitations of DAS, DMAS was suggested in.[33] The same as DAS, DMAS calculates corresponding samples for each element of the array based on the delays, but samples go through a correlation process before adding them up. The DMAS formula is as follows:

$$y_{DMAS}(k) = \sum_{i=1}^{M-1} \sum_{j=i+1}^{M} x_i(k - \Delta_i) x_j(k - \Delta_j). \tag{4}$$

To overcome the dimensionally squared problem of (4), following modifications are suggested in:[33]

$$\hat{x}_{ij}(k) =$$
$$\text{sign}[x_i(k - \Delta_i) x_j(k - \Delta_j)] \sqrt{|x_i(k - \Delta_i) x_j(k - \Delta_j)|}, \tag{5}$$

for $\quad 1 \leqslant i \leqslant j \leqslant M.$

$$y_{DMAS}(k) = \sum_{i=1}^{M-1} \sum_{j=i+1}^{M} \hat{x}_{ij}(k). \tag{6}$$

DMAS algorithm is a correlation process, and a non-linear beamforming algorithm in which the autocorrelation of the aperture is used. A product in time domain is equivalent to the convolution of the spectra of the signals in the frequency domain. Consequently, new components centered at the zero frequency and the harmonic frequency appear in the spectrum due to the similar ranges of frequency for $x_i(k - \Delta_i)$ and $x_j(k - \Delta_j)$. A band-pass filter is applied on the beamformed output signal to only pass the necessary frequency components, generated after the non-linear operations. Having a closer look at (2), the numerator of CF algorithm is the output of DAS beamformer, and the formula can be written as follows:

$$CF(k) = \frac{|y_{DAS}(k)|^2}{M \sum_{i=1}^{M} |x_{id}(k)|^2}. \tag{7}$$

Having CF combined with DAS, (3), leads to sidelobes reduction and contrast enhancement compared to (1). However, in this paper, it is proposed to use the output of DMAS algorithm instead of the DAS algebra on the numerator of CF formula. The proposed weighting is called modified CF (MCF), and its algebra is as follows:

$$MCF(k) = \frac{\left|y_{DMAS}(k)\right|^2}{M \sum_{i=1}^{M} |x_{id}(k)|^2}. \tag{8}$$

MCF will be used the same as CF to weight the samples. The combination of DAS and MCF can be written as follows:

$$y_{DAS+MCF}(k) = MCF(k) \times y_{DAS}(k). \tag{9}$$

Since the DMAS outperforms DAS in the terms of resolution and sidelobes, it is expected that the proposed weighting method provides a higher image quality compared to (2). In what follows, it is shown that the proposed method outperforms the conventional CF weighting.

## 3 Numerical Results and Performance Assessment

In this section, numerical results are presented to evaluate the performance of the proposed algorithm in comparison with DAS and combination of DAS and CF (DAS+CF).

### 3.1 Point Targets

K-wave Matlab toolbox was used to simulate the numerical study.[46] Eleven 0.1 $mm$ spherical absorbers were positioned along the vertical axis every 5 $mm$ as initial pressure. The first absorber was 25 $mm$ away from the transducer surface. The imaging region was 20 $mm$ in lateral axis and 80 $mm$ in vertical axis. A linear array having $M$=128 elements operating at 7 $MHz$ central frequency and 77 % fractional bandwidth was used to detect the PA signals generated from defined initial pressures. The sampling frequency is 50 $MHz$. Speed of sound was assumed to be 1540 $m/s$ during simulations. Envelope detection, performed by means of the Hilbert transform, has been used for all presented images, and the obtained lines are normalized and log-compressed to form the final images.

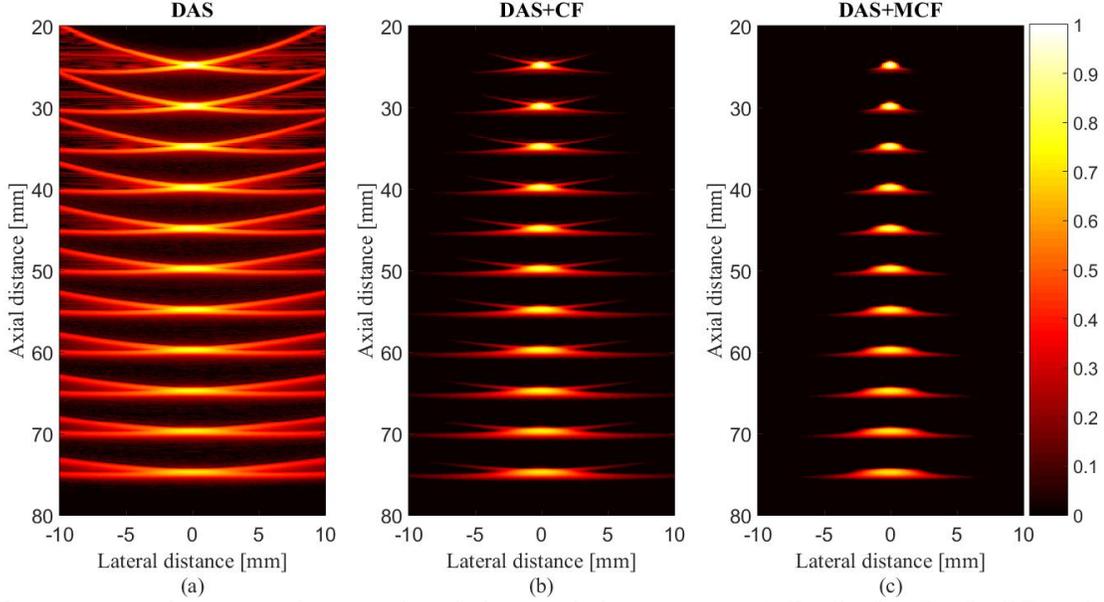

Fig 1: Reconstructed images for simulated detected data using (a) DAS, (b) DAS+CF and (c) DAS+MCF. A linear array and point targets were used for numerical design. All images are shown with a dynamic range of 60 $dB$. Noise was added to the detected signals having a SNR of 50 $dB$.

The reconstructed images are shown in Fig. 1 where Gaussian noise was added to the detected signals having a SNR of 50 $dB$. As is demonstrated, DAS leads to high sidelobes and after the depth of 50 $mm$ the targets are barely detectable as a point target. Using CF combined with DAS results in lower sidelobes and a higher image quality. Fig. 1(c) shows that the proposed method suppresses the artifacts and sidelobes more than the conventional CF. To compare the reconstructed images in detail, the lateral variations at two depths of imaging are shown in Fig. 2. As it is demonstrated, the MCF method causes lower sidelobes. Consider, for instance, the depth of 25 $mm$ where the levels of sidelobes for DAS, DAS+CF and DAS+MCF are for about -36 $dB$, -99 $dB$ and -124 $dB$, which indicates the superiority of the proposed method compared to the conventional CF in the term of sidelobe reduction.

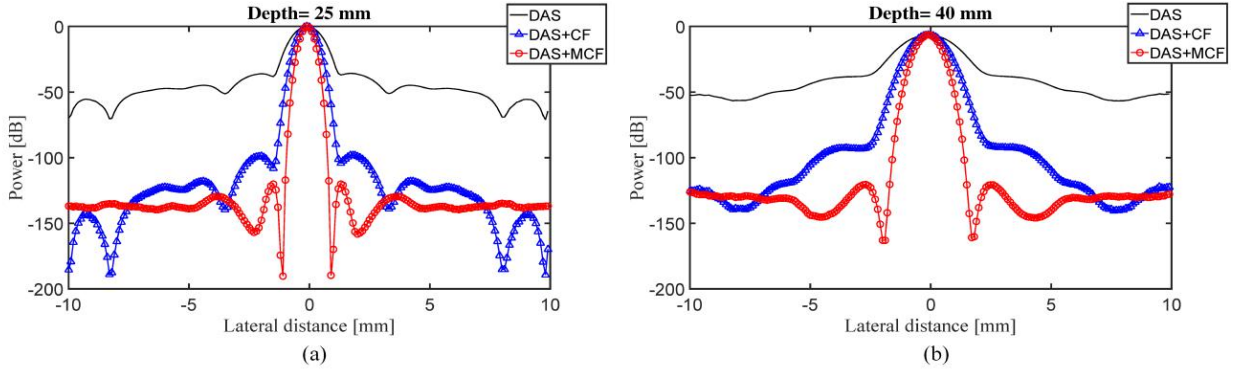

Fig 2: Lateral variations of the reconstructed images shown in Fig. 1 at the depths of (a) 25 $mm$ and (b) 40 $mm$.

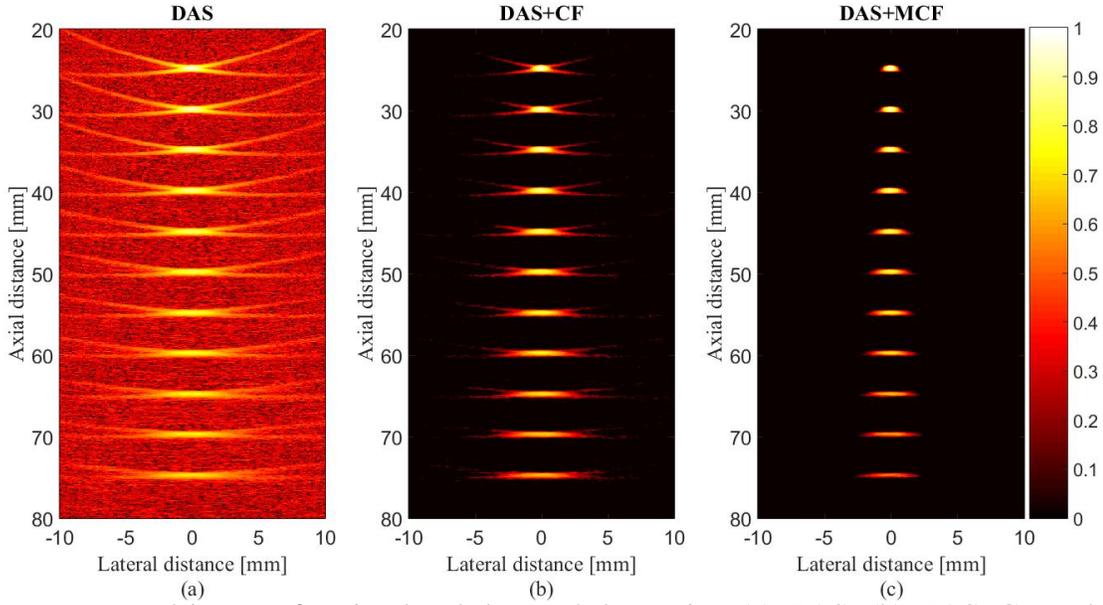

Fig 3: Reconstructed images for simulated detected data using (a) DAS, (b) DAS+CF and (c) DAS+MCF. A linear array and point targets were used for numerical design. All images are shown with a dynamic range of 60 $dB$. Noise was added to the detected signals having a SNR of 0 $dB$.

To evaluate the proposed method at the presence of high level of noise of the imaging system, Gaussian noise was added to the detected signals having a SNR of 0 $dB$. The reconstructed images are shown in Fig. 3, and as can be seen, the formed image obtained by DAS is highly affected by noise. CF improves the image quality by suppressing the effects of noise. However, the sidelobes still degrade the image quality. The MCF reduces the sidelobes and improves the target detectability, resulting in a higher image quality in comparison with CF. It should be noticed that the absence of the tails attached to the targets (can be seen in Fig. 1) is due to the high level of noise. To put it more simply, the power of noise is more than the tails, and that is why they are not seen in Fig. 3. The lateral variations for the images shown in Fig. 3, are shown in Fig. 4, and as can be seen, the higher performance of MCF in the terms of sideloebs and noise suppression, compared to CF, is clear.

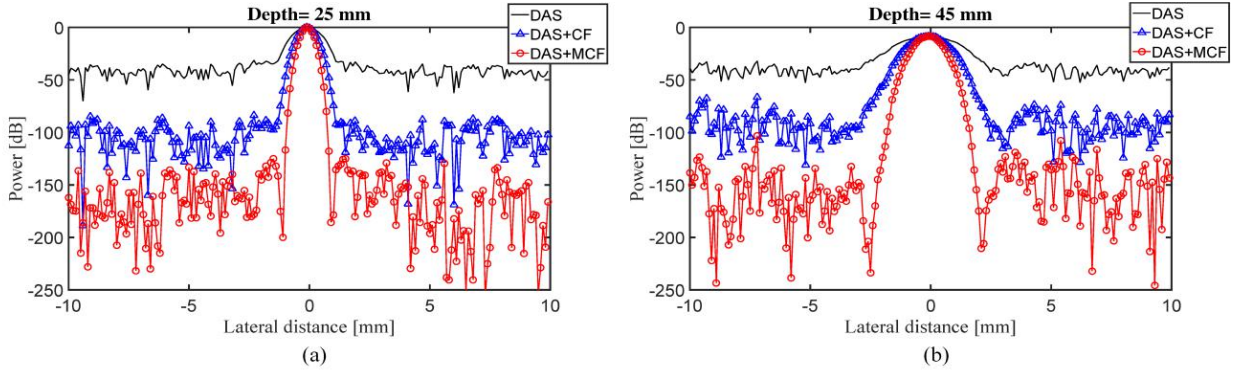

Fig 4: Lateral variations of the reconstructed images shown in Fig. 3 at the depths of (a) 25 $mm$ and (b) 45 $mm$.

*3.2 Quantitative Evaluation*

To quantitatively assess the performance of the proposed weighting method, the full-with-half-maximum (FWHM) in -6 $dB$ and signal-to-noise ratio (SNR) are calculated and presented in Table 1 and Table 2, respectively. SNR is calculated using the method explained in.[37] As shown in Table 1, the FWHM gained by MCF, at the all depths, is lower than CF, showing the superiority of MCF. Consider, for instance, the depth of 40 $mm$ where DAS, DAS+CF and DAS+MCF results in 2.2 $mm$, 1.3 $mm$ and 0.9 $mm$, respectively. In other word, MCF improves the FWHM for about 0.4 $mm$ compared to CF. As shown in Table 2, SNR gained by the proposed weighting method is higher compared to the CF which also indicates the superiority of MCF. Consider, for example, the depth of 50 $mm$ where the SNR for DAS, DAS+CF and DAS+MCF is for about 36.9 $dB$, 65.0

Table 1: -6 $dB$ FWHM ($mm$) values at the different depths.

| Beamformer<br>Depth($mm$) | DAS | DAS+CF | DAS+MCF |
|---|---|---|---|
| 25 | 1.1 | 0.6 | 0.4 |
| 30 | 1.3 | 0.8 | 0.6 |
| 35 | 1.6 | 0.9 | 0.7 |
| 40 | 1.9 | 1.1 | 0.8 |
| 45 | 2.2 | 1.3 | 0.9 |
| 50 | 2.6 | 1.6 | 1.1 |
| 55 | 3.0 | 1.8 | 1.3 |
| 60 | 3.5 | 2.1 | 1.5 |
| 65 | 3.7 | 2.2 | 1.6 |
| 70 | 4.2 | 2.5 | 1.8 |
| 75 | 4.8 | 2.9 | 2.0 |

Table 2: SNR ($dB$) values at the different depths.

| Beamformer<br>Depth($mm$) | DAS | DAS+CF | DAS+MCF |
|---|---|---|---|
| 25 | 47.2 | 76.3 | 119.6 |
| 30 | 44.7 | 73.0 | 116.9 |
| 35 | 43.0 | 72.9 | 117.8 |
| 40 | 40.7 | 69.5 | 116.5 |
| 45 | 38.9 | 68.0 | 113.3 |
| 50 | 36.9 | 65.0 | 110.4 |
| 55 | 35.3 | 63.4 | 109.0 |
| 60 | 34.2 | 61.8 | 107.3 |
| 65 | 33.5 | 60.1 | 105.3 |
| 70 | 32.2 | 58.6 | 103.2 |
| 75 | 31.4 | 56.2 | 101.6 |

$dB$ and 110.4 $dB$, respectively.

*3.3 MCF Applied to DMAS*

It should be noted that the proposed method in this paper is a weighting technique which can be applied to any beamformer to achieve a higher image quality. Here, the aim is to evaluate the MCF when is applied to other beamformers except DAS. The DMAS beamformer was selected. The results are presented in Fig. 5. As demonstrated, the proposed method leads to higher sidelobes reduction and artifacts removal when it is applied on the DMAS, compared to CF. In other words, even though the CF degrades the sidelobes (in DAS and DMAS), MCF outperforms the conventional CF with a higher artifacts suppression. For further evaluation, consider the lateral variations at the depth of 45 $mm$, shown in Fig. 6, where the MCF reduces the sidelobes for about 20 $dB$, compared to the conventional CF.

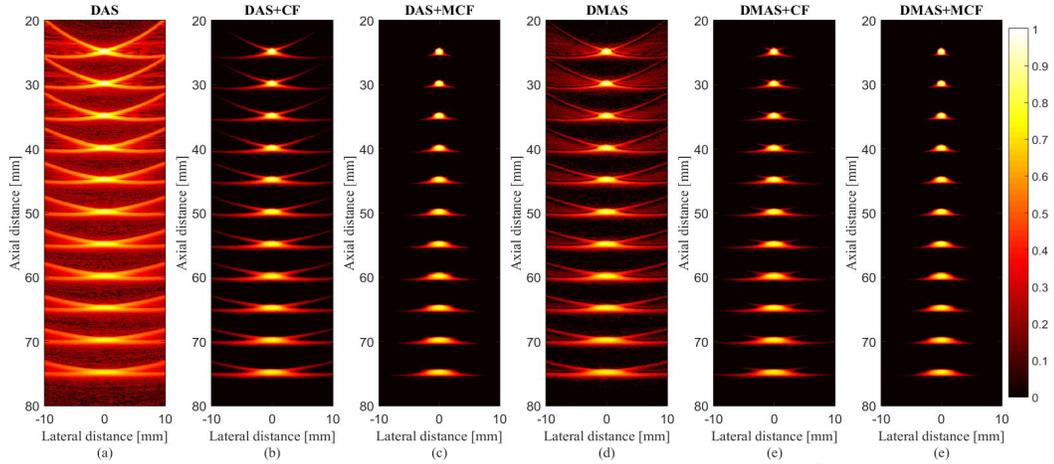

Fig 5: Reconstructed images for simulated detected data using (a) DAS, (b) DAS+CF, (c) DAS+MCF, (d) DMAS, (e) DMAS+CF and (e) DMAS+MCF. A linear array and point targets were used for numerical design. All images are shown with a dynamic range of 70 $dB$. Noise was added to the detected signals having a SNR of 20 $dB$.

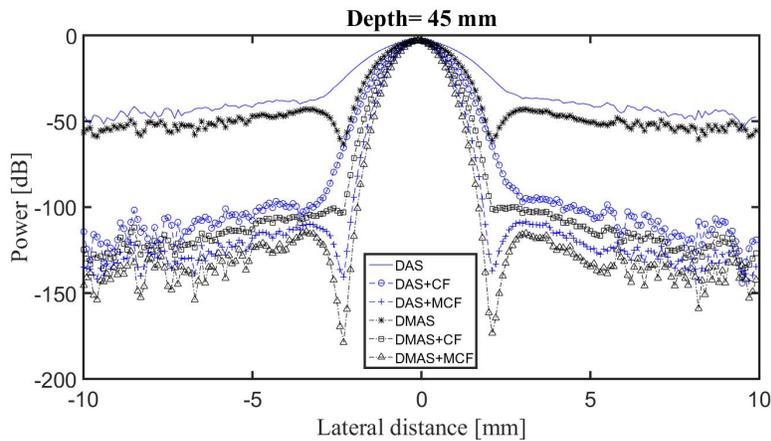

Fig 6: Lateral variations of the reconstructed images shown in Fig. 5 at the depths of 45 $mm$.

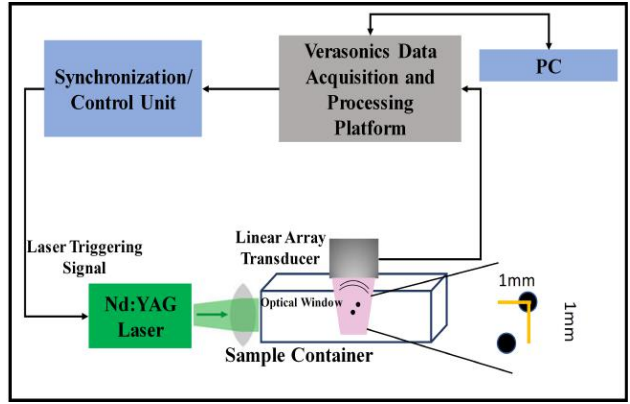

Fig 7: The schematic of the setup used for the experimental PAI.

Table 3: FWHM ($mm$) values, in -6 $dB$, at the two depths of imaging using the experimental data.

| Beamformer<br>Depth($mm$) | DAS | DAS+CF | DAS+MCF |
|---|---|---|---|
| 22 | 0.68 | 0.59 | 0.30 |
| 24 | 0.66 | 0.47 | 0.28 |

## 4 Experimental Results

To further evaluate the proposed weighting method and its effect on enhancing PA images, phantom experiments were performed in which a phantom consists of 2 light absorbing wires with diameter of 150 $\mu m$ were placed 1 $mm$ apart from each other in a water tank. The schematic of the experimental setup is shown in Fig. 7. In this experiment, we utilized a Nd:YAG pulsed laser, with the pulse repetition rate of 30 $Hz$ at wavelengths of 532 $nm$. A programmable digital ultrasound scanner (Verasonics Vantage 128), equipped with a linear array transducer (L11-4v) operating at frequency range between 4 to 9 $MHz$ was utilized to acquire the PA RF data. A high speed FPGA was used to synchronize the light excitation and PA signal acquisition.

The reconstructed images are shown in Fig. 8. As it is demonstrated, DAS results in high levels of noise in the images, degrading the image quality, and image is affected by sidelobes. Using CF improves the images quality, but the image is still affected by noise and sidelobes. Finally, the proposed weighting method enhances the image by providing higher noise suppression and lower levels of sidelobes compared to conventional CF. To evaluate in more detail, the lateral variations at two depths for targets shown in Fig. 8, are presented in Fig. 9. Considering Fig. 9(a), the proposed weighting method results in lower sidelobes and noise where DAS, DAS+CF and DAS+MCF leads to -40 $dB$, -84 $dB$ and -134 $dB$, respectively. Thus, the proposed method outperforms conventional CF. FWHM in -6 $dB$ has been calculated for the experimental results and shown in Table 3. It can be seen that the proposed weighting method results in narrower mainlobe in comparison with CF. Consider, for example, the target at the depth of 22 $mm$ where DAS+MCF results in 0.38 $mm$ 0.29 $mm$ improvement compared to DAS and DAS+CF, respectively. Moreover, SNR has been calculated for the experimental data and the results are shown in Table 4 where the MCF causes higher SNR compared to conventional CF for both depths of imaging.

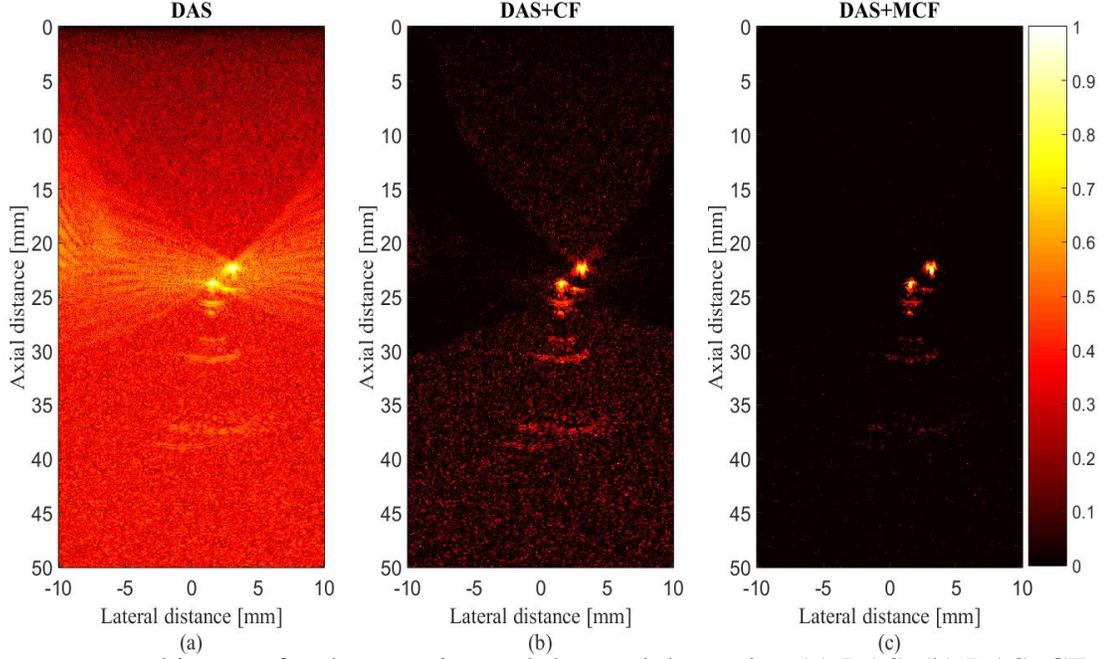

Fig 8: Reconstructed images for the experimental detected data using (a) DAS, (b) DAS+CF and (c) DAS+MCF. A linear array and wire target phantom were used for the experimental design. All images are shown with a dynamic range of 80 $dB$.

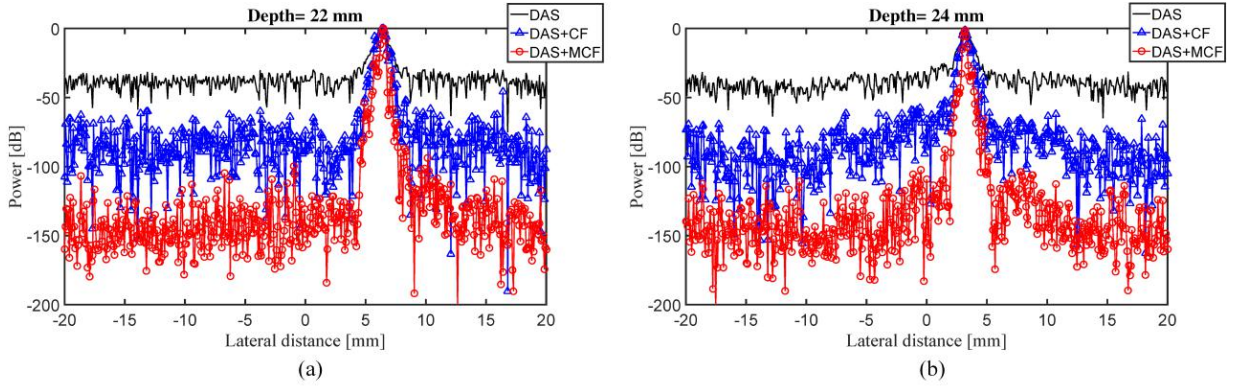

Fig 9: Lateral variations of the reconstructed images shown in Fig. 8 at the depths of (a) 22 $mm$ and (b) 24 $mm$.

Table 4: SNR ($dB$) values for the experimental images shown in Fig. 8.

| Beamformer<br>Depth($mm$) | DAS | DAS+CF | DAS+MCF |
|---|---|---|---|
| 22 | 48.5 | 60.7 | 90.4 |
| 24 | 47.2 | 59.6 | 88.2 |

*4.1 Ex Vivo Imaging*

In this study, an *ex vivo* experimental tissue study have been designed to evaluate the performance of the proposed algorithm. A piece of a breast tissue (about 4 $cm\times$ 4 $cm\times$ 3 $cm$) is extracted from a new sacrificed chicken. Two pencil leads with a diameter of 0.5 $mm$ are embedded inside the breast tissue, having an axial distance of about 5 $mm$. Fig 10 shows the photographs of the imaged tissue. The PA signals are collected with a combined linear US/PA imaging probe.[47] As can be seen in Fig. 11(a), the artifacts and the background noise degrade the PA image quality obtained by DAS. As is expected based on the previous results, applying CF to the DAS algorithm would reduce the artifacts and sidelobes. The expectations are satisfied, as shown in Fig. 11(b), but the PA image can be further improved using MCF. As demonstrated in Fig. 11(c), the MCF leads to higher noise suppression and sidelobes degrading in comparison with the conventional CF. For further evaluation, the lateral variations of the reconstructed images shown in Fig. 11, are presented in Fig. 12 where the superiority of the proposed method in terms of lower sidelobes and higher noise suppression is obvious. SNR is calculated for the *ex vivo* images (presented in Table 5). The quantitative evaluation indicates that MCF outperforms the conventional CF. In particular, it improves the SNR for about 14 $dB$, at the depth of 31.3 $mm$, compared to the CF.

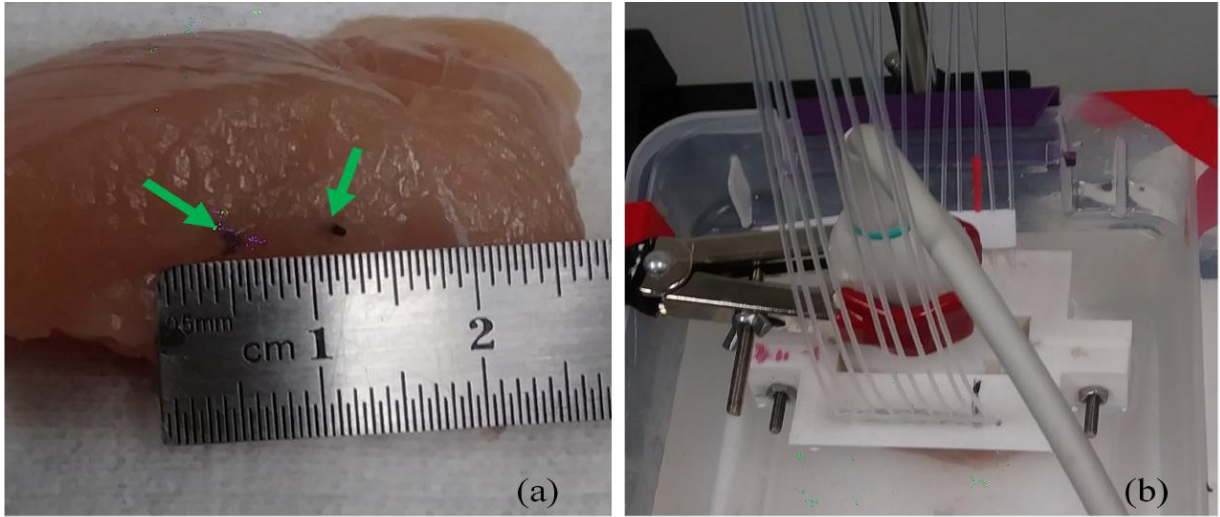

Fig 10: (a) The phantom used for the experiment. (b) The *ex vivo* imaging setup.

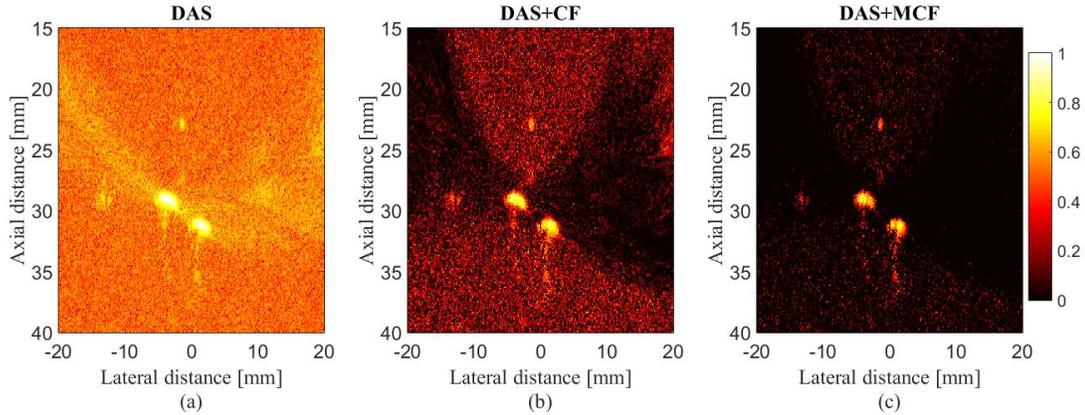

Fig 11: Reconstructed *ex vivo* images using (a) DAS, (b) DAS+CF and (c) DAS+MCF. A linear-array and the phantom shown in Fig. 10 were used for the experimental design. All images are shown with a dynamic range of 80 $dB$.

## 5 Discussion

The main enhancement gained by the proposed method is higher contrast and lower sidelobes. Considering the fact that DAS beamformer results in a low quality image, having it on the numerator of the formula of CF would degrade the performance of the CF weighting procedure. On the other hand, in,[33] Matrone *et al.* proved that DMAS can be used instead of DAS for image reconstruction, and it was shown that the main improvement gained by DMAS was higher contrast. Thus, it can be perceived that using DMAS instead of the existing DAS on the numerator of CF algebra would results in contrast enhancement due to its auto-correlation process which is a non-linear operation. As can be seen in Fig. 1, Fig. 3, Fig. 8 and Fig. 11, using the correlation process of DMAS inside the formula of CF results in higher noise suppression and artifact reduction, leading to the higher image quality compared to DAS and DAS+CF. In other word, the multiplication operation inside the DMAS procedure reduces the presence of noise and off-axis signals on the reconstructed images, and improves the image quality. The advantage of the proposed weighting method in the term of sidelobes reduction can be seen in the Fig. 2, Fig. 4, Fig. 9 and Fig. 12. As can be seen in the lateral variations, the width of mainlobe has decreased which is a merit of MCF. Since DMAS improves the resolution gained by DAS, shown in,[37] the MCF leads to higher resolution in comparison with CF. To put it more simply, presence of DMAS inside the formula of MCF is the reason of higher resolution achieved by MCF. Despite all the results, it was necessary to evaluate the proposed method quantitatively. Considering the numbers presented in the Tables of the last section, it can be seen that the proposed method outperforms CF in the terms of FWHM and SNR. The proposed method significantly outperforms CF when the targets are at the high depths of imaging. As shown in Fig. 1, for the targets located at the depths of 55 $mm$-75 $mm$, sidelobes and artifatcs are better reduced compared to the lower depths. This also can be perceived regarding the Table 2 where SNR improvement in high depths is more than lower depths (80 % and 56 % for 25 $mm$ and 75 $mm$, respectively). It should be noticed that, as mentioned in the section 1, the beamforming and concerned weighting methods can be applied on both the US and PAI cases. CF or MCF would be proper options for US imaging. However, using multiple times of the CF or MCF ( $MCF^2$ and $MCF^3$) would remove the speckles in the US images. The speckle removal is not desired in applications in which the speckles provide helpful information for diagnosis. Therefore, the $MCF^2$ and $MCF^3$ would further increase the quality of the PA images, but it is not suggested to use them for the conventional US imaging. MCF The computational burden imposed by the proposed method is the same as DMAS, and the order of processing in O($M^2$) while the order of processing for CF is the same as DAS which is O($M$). Therefore, it should be mentioned that the improvements are obtained at the expense of higher computational burden in comparison with CF. The proposed method can be implemented on a FPGA device, e.g. on an Altera FPGA of the Stratix IV family (Altera Corp., San Jose, CA, USA). The time consumption has been reported in[33] for DMAS implementation, which indicates that the proposed method can be used in clinical PAI systems. In applications in which phased (or micro-convex) arrays are used, MCF can provide a further enhancement compared to CF. We have tested the proposed algorithm for *ex vivo* PAI, and the results were promising (shown in Fig. 11 and Fig. 12). In small-parts and vascular US imaging, for instance *in vivo* imaging of the carotid artery, where the resolution and specially sidelobes are of importance, MCF can be used, providing higher contrast and noise suppression in comparison with CF.

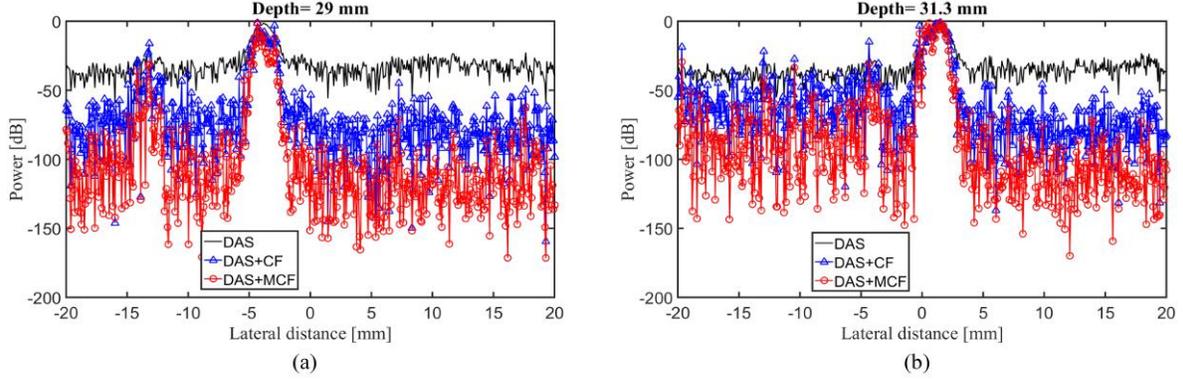

Fig 12: Lateral variations of the reconstructed images shown in Fig. 11 at the depths of (a) 29 $mm$ and (b) 31.3 $mm$.

Table 5: SNR ($dB$) values for the *ex vivo* images shown in Fig. 11.

| Beamformer<br>Depth($mm$) | DAS | DAS+CF | DAS+MCF |
|---|---|---|---|
| 29 | 41.34 | 52.81 | 68.99 |
| 31.3 | 40.35 | 50.78 | 64.84 |

# 6 Conclusion

In this paper, a novel weighting procedure has been introduced by combining the conventional CF and the DMAS beamformer. It was shown that the numerator of the formula of CF can be treated as a DAS, and it was proposed to use DMAS instead of the existing DAS inside the formula of CF. The MCF has been evaluated numerically and experimentally, and all the results showed the higher performance of MCF compared to CF. For the experimental results obtained by the wire target phantom, MCF reduced the sidelobes for about 50 $dB$ in comparison with CF, indicating the higher contrast, and the quantitative results showed that MCF improves the SNR and FWHM for about 45% and 40%, respectively.

*Acknowledgments*

This research received no specific grant from any funding agency in the public, commercial, or not-for-profit sectors, and the authors have no potential conflicts of interest to disclose.

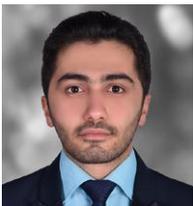**Moein Mozaffarzadeh** was born in Sari, Iran, in 1993. He received the B.Sc. degree in Electrical Engineering from Babol Noshirvani University of Technology (Mazandaran, Iran), in 2015, and the M.Sc. degree in Biomedical Engineering from Tarbiat Modares University (Tehran, Iran), in 2017. He is currently a research assistant at research center for biomedical

technologies and robotics, institute for advanced medical technologies (Tehran, Iran). His current research interests include Photoacoustic Image Reconstruction, Ultrasound Beamforming and Biomedical Imaging.

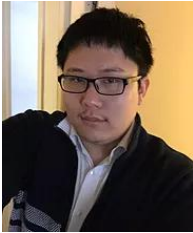 **Yan Yan** joined Wayne State University since 2015-May as a PhD student in functional and molecular ultrasound research laboratory (http://ultrasound.eng.wayne.edu/). His interesting areas are Medical Image, Object Detection and Data Mining. He had strong background in Computer Science pattern recognition and computer graphics. He holds two bachelors, Computer science and Post and Telecommunication. He also received a master in computer science from Wayne State university 2017. His current research interests include Endocavity Ultrasound and Photoacoustic for fetal and maternal care.

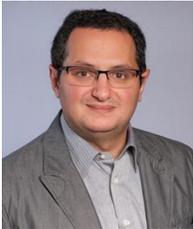 **Mohammad Mehrmohammadi** received his B.Sc. degree in Electrical Engineering from Sharif University of Technology (Tehran, Iran), the M.Sc. in Electrical and Computer Engineering from Illinois Institute of Technology (Chicago, IL), and the Ph.D. in Biomedical Engineering from the University of Texas at Austin (Austin, TX). He did his postdoctoral fellowship at Mayo Clinic College of Medicine (Rochester, Minnesota). Currently, he is an assistant professor of Biomedical/Electrical and Computer Engineering at Wayne State University and scientific member at Karmanos Cancer Institute.

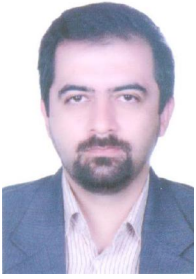 **Bahador Makkiabadi** received his B.Sc. degree in Electronics Engineering from Shiraz University (Shiraz, Iran), the M.Sc. in Biomedical Engineering from Amirkabir University of Technology (Tehran, Iran), and the Ph.D. in Biomedical Engineering from University of Surrey (Guildford, Surrey, UK). Currently, he is working at Research Center for Biomedical Technologies and Robotics (RCBTR), Institute for Advanced Medical Technologies (IAMT), Tehran University of Medical Sciences, Tehran, Iran. His research interests include Blind Source Separation, Advanced Array Signal Processing for Medical Applications and Biomedical Imaging.

## List of Figures

## List of Tables

2  SNR ($dB$) values at the different depths.

3  FWHM ($mm$) values, in -6 $dB$, at the two depths of imaging using the experimental data.

4  SNR ($dB$) values for the experimental images shown in Fig. 8.

5  SNR ($dB$) values for the *ex vivo* images shown in Fig. 11.